\documentclass[11pt]{article}

\usepackage{graphicx}
\usepackage{amsmath,amssymb,amsthm}
\usepackage{asymptote}

\title{The Snowblower Problem: An Improvement}
\author{Justine Tang}
\date{}

\begin{document}
\maketitle

\begin{abstract}
We present more optimal solutions to the snowblower problem introduced in \cite{snow}. In particular, we present more optimal ways to clear lines and combs, which are shapes as described in the aforementioned paper that the original input is dissected into. We note that our methods to clear lines and combs reduce the number of steps needed to clear such shapes, while also acknowledging that they do not change the overall complexity.
\end{abstract}

\section{Introduction}

The snowblower problem (SBP), as introduced in \cite{snow}, is a problem intended to model a real-world snowblower. In the SBP, a snowblower is tasked with moving all the snow in a driveway, modeled as a rectilinear polygon, to the edge of the driveway, with a few constraints. For one, every pixel on the driveway is assumed to be covered with exactly $1$ unit of snow. Additionally, no pixel on the driveway can have a depth of more than $D$ units of snow, where $D$ is an integer greater than $1.$

As stated in \cite{snow}, the SBP is NP-complete. As such, the solution given in that paper, as well as the solution presented here, are not optimal, but rather approximate the optimal solution to the problem. The paper's proposed solution involves dissecting the input polygon into Voronoi cells, such that for each pixel in the Voronoi cell of side $e,$ $e$ is the closest side to that pixel. It can be proven that all Voronoi cells are either lines, or a straight row or column of pixels; combs, or a collection of straight rows or columns of pixels, with either all the leftmost or rightmost pixels in the same column or row; or double-sided combs, or two opposite-oriented combs joined together by their ``handle'' or common column or row. Further specifics about lines and combs can be found in the original paper.

The original paper proposed various methods to clear lines, combs, and double-sided combs; however, it can be proven that the following methods, while preserving the overall complexity of the algorithm, require fewer steps to carry out. We use much of the same terminology for lines and combs as that defined in \cite{snow}, such as ``root'', ``tooth'', and ``handle''.

\section{Improving Line Clearing}

We first propose a method to clear a line that requires less steps than that given in \cite{snow}.

Let the line have a length of $L;$ initially, there are $L$ units of snow in the line. Let $R=L\,\%\,D$ and $Q=\lfloor\frac{L}{D}\rfloor,$ so that $L=R+QD,$ and that $R$ is the remainder and $Q$ the quotient of $L$ divided by $D.$ We start at the root of the line.

We can first clear $R$ units of snow from the line so that the number of units remaining is divisible by $D.$ In the below diagrams, $L=10$ and $D=4.$

\begin{center}
\includegraphics[scale=0.4]{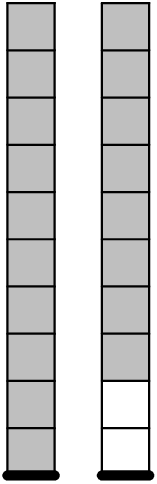}
\end{center}

We make $2(R-1)$ steps in the process, as we make $R-1$ steps both back and forth.

Then, we can continuously make $D$-full passes until all of the snow is cleared. We make $2(R+D-1)$ steps in the first pass, $2(R+2D-1)$ in the second, and so on, until we reach $2(L-1)=2(R+QD-1)$ steps on the last pass.

\begin{center}
\includegraphics[scale=0.4]{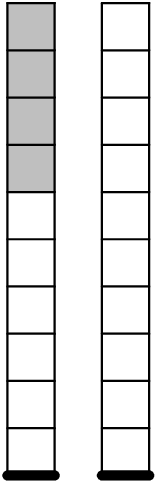}
\end{center}

This gives us a total of
$$2[(R-1)+(R+D-1)+\dots+(R+QD-1)]$$
steps, which simplifies to $(Q+1)(2R+QD-2)$ steps.

In contrast, \cite{snow} proposes to continuously make $D$-full passes before clearing the remainder. We make $2(D-1)$ steps in the first pass, $2(2D-1)$ in the second, up until $2(QD-1)$ steps in the $Q$th pass; finally, we make $2(L-1)=2(R+QD-1)$ steps in the last pass to clear the rest of the line.

\begin{center}
\includegraphics[scale=0.4]{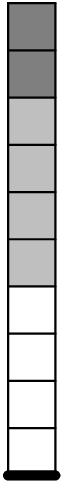}
\end{center}

This method gives us a total of
$$2[(D-1)+(2D-1)+\dots+(QD-1)+(R+QD-1)]$$
steps, which simplifies to $Q[(Q+1)D-2]+2(R+QD-1)$ steps.

Expanding both of these expressions, we find that our proposed method yields
$$2QR+2R+Q^2D+QD-2Q-2$$
steps, while the original paper's method yields
$$Q^2D+QD-2Q+2R+2QD-2$$
steps. We see that these two expressions are the same except for one term: our proposed method yields a $2QR$ term, while the paper's method yields a $2QD$ term. Since $R=L\%D,$ we know that $R<D,$ so $2QR<2QD$ and our method saves a total of
$$2QD-2QR=2Q(D-R)$$
steps from the paper's method.

We note that the exception to this is when $R=0;$ then both methods lead to making continuous $D$-full passes until the line is clear. This means that both methods give a total of
$$2[(D-1)+(2D-1)+\dots+(QD-1)]$$
steps, which simplifies to $Q[(Q+1)D-2]$ steps; no steps are saved (or gained) using our method.

\section{Improving Comb Clearing}

We now propose a method to clear a comb that requires less steps than that given in \cite{snow}. We first consider that of a single-sided comb. As described in the paper, let $p$ be the root of the comb, and let the handle have length $H.$ Additionally, let $T_i$ be the length of tooth $i,$ including the pixel at the handle, where $T_0$ is the first tooth. We also define $R_i$ as $T_i\,\%\,D$ if $D$ does not divide $T_i$ (i.e., if $T_i\,\%\,D>0$), and $D$ otherwise. We also define $Q_i$ as the integer satisfying $T_i=R_i+Q_iD,$ resembling the terminology used in the previous section.

To begin, we can first perform a ``brush'' on the comb like that described in \cite{snow}. However, instead of brushing the entire comb, we only brush the ``remainders'' of each tooth; i.e., we only brush the first $R_i$ pixels of each tooth. Additionally, in order to prevent more than $D$ pixels from being on any given pixel, we start our depth-first search from the first uncleared tooth, rather than the last one. An example of this is given below, with $D=4;$ the shaded pixels represent the pixels to be brushed, and the path shown represents the first pass of the brush.

\begin{center}
\includegraphics[scale=0.4]{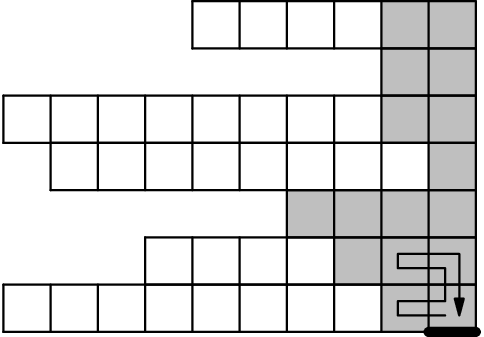}
\end{center}

After this, we can begin making $D$-full passes on each tooth, since the snow remaining on each tooth is now divisible by $D.$ For each tooth $i,$ we can now move through the first $R_i$ pixels and make $D$-full passes on the tooth starting from pixel $R_i+1$ until there is no snow remaining past that pixel. An example of this is shown below, where we take the first tooth into consideration.

\begin{center}
\includegraphics[scale=0.4]{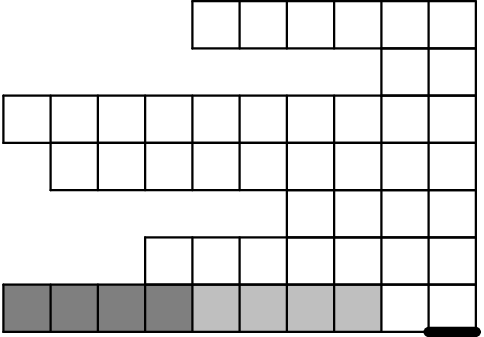}
\end{center}

In order to find the difference in number of steps between our method and the method described in \cite{snow}, we can first count the number of steps made via $D$-full passes on each tooth. For brevity, we refer to the method described in \cite{snow} as simply ``the paper's method''.

Using our method, we know that when we make the $D$-full passes, we skip the first $R_i$ pixels of tooth $i,$ moving to pixel $R_i+D$ and back to the handle, and then to $p,$ to make our first pass. Then we can move back to the tooth and move to pixel $R_i+2D$ and back, and so on until pixel $R_i+Q_iD.$ We can see it takes $2i$ steps to move to tooth $i$ and back, as well as $2(R_i+jD-1)$ steps to move from the handle of tooth $i$ to the desired pixel, where $j$ is the number of passes that have been made, including the current one.

Thus, for tooth $i,$ the $D$-full passes yield a total of
$$\sum^{Q_i}_{j=1}2(R_i+jD-1+i)$$
steps, and iterating on the comb as a whole, the passes yield a total of
$$\sum^{H-1}_{i=0}\sum^{Q_i}_{j=1}2(R_i+jD-1+i)$$
steps.

The paper's method for $D$-full passes is largely the same; however, since it does not skip the first $R_i$ pixels of the tooth, we can simply leave the $R_i$ term of the above summation out. This gives us a total of
$$\sum^{H-1}_{i=0}\sum^{Q_i}_{j=1}2(jD-1+i)$$
steps from the paper's method. Subtracting these two expressions, we find that our method yields
$$\sum^{H-1}_{i=0}\sum^{Q_i}_{j=1}2R_i=2\sum^{H-1}_{i=0}Q_iR_i$$
more steps than the paper's method, when only taking $D$-full passes into account.

We can now find the difference in number of steps when considering the brush. Using the paper's method, all of the remaining snow will be at the tip of each tooth; this means that for each pass through the comb we make during the brush, we will have to traverse the entire tooth and back. On the other hand, using our method, all the remaining snow will be at the handle of each tooth, so we will only have to traverse the first $R_i$ pixels.

\begin{center}
\includegraphics[scale=0.4]{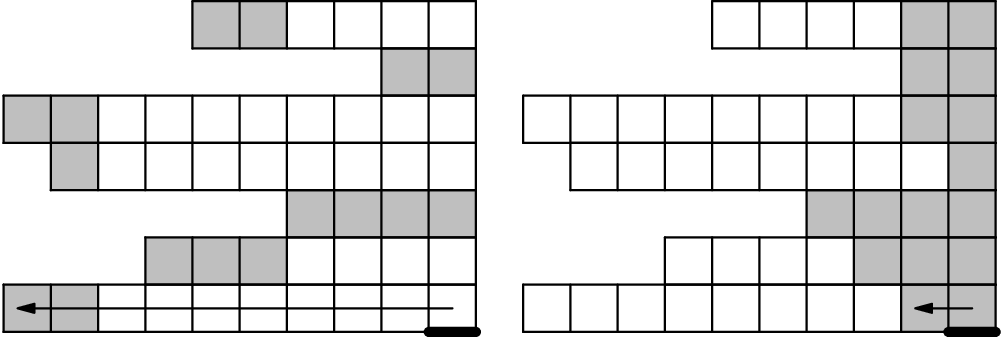}
\end{center}

This means that in total, the paper's method has to make $2(T_i-R_i)$ steps more than our method when moving through and clearing a tooth. We know that we have to move through every tooth at least once, in order to clear all the snow from that tooth. Due to this, we find that our method saves at least
$$\sum^{H-1}_{i=0}2(T_i-R_i)=2\sum^{H-1}_{i=0}Q_iD$$
steps from the paper's method.

Therefore, we find that in total, our method saves at least
$$2\sum^{H-1}_{i=0}(Q_iD-Q_iR_i)=2\sum^{H-1}_{i=0}Q_i(D-R_i)$$
steps from the paper's method. Since $R_i\leq D$ by definition, we know that this summation is nonnegative.

This method can be generalized to double-sided combs; one way of approaching this would be to split the double-sided comb into two single-sided combs like below, where the shaded part represents the comb that is being ``broken off'' from the original double-sided comb.

\begin{center}
\includegraphics[scale=0.4]{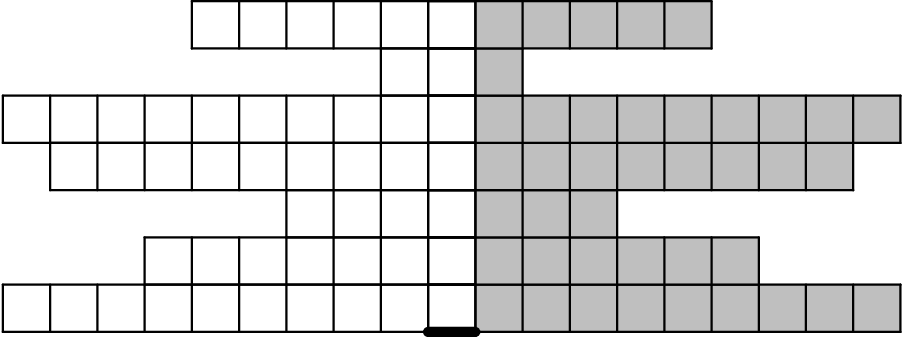}
\end{center}

\bibliography{references}
\bibliographystyle{ieeetr}
\end{document}